\documentclass[journal=acsnano,manuscript=article]{achemso}
\usepackage{graphicx,color}
\usepackage{dcolumn}
\usepackage{bm}

\def\k{\kappa}

\def\w{\omega}

\def\bk{{\bf k}}

\def\bG{{\bf G}}
\def\bq{{\bf q} }

\def\bR{{\bf R} }
\def\bQ{{\bf Q} }
\def\bt{{\bm \tau} }

\def\ve{\varepsilon}

\def\<{\langle}
\def\>{\rangle}
\def\D{\partial}
\let\hide\iffalse

\usepackage{braket} 
\usepackage{colortbl}
\usepackage[final]{changes}
\usepackage{amsmath}

\author{H\'el\`ene Seiler}
\email{seiler@fhi-berlin.mpg.de}
\affiliation{Fritz Haber Institute of the Max Planck Society, 14195 Berlin, Germany}
\author{Daniela Zahn}
\affiliation{Fritz Haber Institute of the Max Planck Society, 14195 Berlin, Germany}
\author{Marios Zacharias}
\affiliation{Fritz Haber Institute of the Max Planck Society, 14195 Berlin, Germany}
\alsoaffiliation{Department of Mechanical and Materials Science
  Engineering, Cyprus University of Technology, P.O. Box 50329, 3603
  Limassol, Cyprus}
\author{Patrick Hildebrandt}
\affiliation{Fritz Haber Institute of the Max Planck Society, 14195 Berlin, Germany}
\author{Thomas Vasileiadis}
\affiliation{Fritz Haber Institute of the Max Planck Society, 14195 Berlin, Germany}
\author{Yoav William Windsor}
\author{Yingpeng Qi}
\affiliation{Fritz Haber Institute of the Max Planck Society, 14195 Berlin, Germany}
\author{Christian Carbogno}
\affiliation{Fritz Haber Institute of the Max Planck Society, 14195 Berlin, Germany}
\author{Claudia Draxl}
\affiliation{Institut f\"ur Physik and IRIS Adlershof, Humboldt-Universit\"at zu Berlin, Berlin, Germany}
\author{Ralph Ernstorfer}

\affiliation{Fritz Haber Institute of the Max Planck Society, 14195 Berlin, Germany}
\author{Fabio Caruso}
\affiliation{Institut f\"ur Theoretische Physik und Astrophysik, Christian-Albrechts-Universit\"at zu Kiel, D-24098 Kiel, Germany}
\email{caruso@physik.uni-kiel.de}

\title{Accessing the anisotropic non-thermal phonon populations in black phosphorus}
\abbreviations{FEDS}
\keywords{Femtosecond electron diffraction, 2D materials, momentum-resolved phonon dynamics, DFT}

\begin{document}

\begin{abstract}
We combine femtosecond
electron diffuse scattering experiments and first-principles calculations of the coupled
electron-phonon dynamics to provide a detailed momentum-resolved picture of the
ultrafast lattice thermalization in a thin film of black phosphorus.
The measurements reveal the emergence of highly anisotropic non-thermal phonon populations
which persist for several picoseconds following excitation of the electrons with a light pulse.
Combining ultrafast dynamics simulations based on the time-dependent Boltzmann formalism
and calculations of the structure factor, we reproduce the experimental data and
identify the vibrational modes primarily responsible for the carrier relaxation
via electron-phonon coupling and the subsequent lattice thermalization via phonon-phonon scattering.
In particular, we attribute the non-equilibrium lattice dynamics of black phosphorus
to highly-anisotropic phonon-assisted scattering processes, which are primarily
mediated by high-energy optical phonons. Our approach paves the way towards unravelling and controlling microscopic energy-flow pathways in two-dimensional materials and van der Waals heterostructures, and may also be extended to other non-equilibrium phenomena involving coupled electron-phonon dynamics such as superconductivity, phase transitions or polaron physics. 
\end{abstract}

\maketitle
Black phosphorus (BP) exhibits a tunable band gap in the mid-IR
\cite{2014Qiao,2015Cast, 2016Li}, high carrier mobilities
\cite{2014Li,2014Xia,2016Long}, and a layered crystal structure.
These features make it a versatile platform to explore novel device concepts, such as
field-effect transistors, saturable absorbers, and polarization-sensitive
photodetectors \cite{2014Li, 2014Xia, Ling2015, 2015Cast, Buscema2014, Sotor2015}.  The
pronounced crystal structure anisotropy of BP further underpins the emergence of
strikingly anisotropic macroscopic properties, as exemplified by
its thermal \cite{2015Lee,2015Luo,2015Jang} and electrical conductivities \cite{2014Qiao,2014Xia,2014Liu,He2015},
as well as its optical response \cite{2014Xia,2014Tran,2014Low,Jiang2018}.

Since practical applications based on these properties invariably exploit
non-equilibrium states of either the lattice or hot carriers,
it is desirable to attain a detailed understanding of the ultrafast dynamics of electronic
and vibrational degrees of freedom in BP.
Following photo-excitation, hot carriers relax to the band edges by
transferring their excess energy to the lattice via the emission of phonons, which
triggers coupled carrier-lattice non-equilibrium dynamics.
Optical and photoemission spectroscopies have been employed extensively to investigate
carrier-phonon scattering channels and their influence on the carrier dynamics in
BP \cite{He2015, Ge2015, Suess2015,Wang2016,Iyer2017,Liao2017,Meng2019, Roth2019, Chen2019}.
While these techniques provide direct information on the electrons,
the non-equilibrium dynamics of the lattice can only be inferred indirectly through its
effects on the electronic structure. Femtosecond electron diffuse scattering (FEDS), conversely, is an ideal technique to circumvent these limitations and complement optical and photoemission spectroscopies. FEDS provides direct access to lattice dynamics and electron-phonon scattering processes
with time and momentum resolution \cite{Stern2018,Cotret2019,Waldecker2017}.
Owing to its sensitivity to both electron-phonon and
phonon-phonon scattering processes in reciprocal space,
FEDS is thus well-suited to establish a microscopic picture
of the energy flow between hot electrons and the BP lattice.

Here, we combine FEDS experiments with ab-initio calculations
to investigate the coupled electron-phonon dynamics in BP following photo-excitation. Our study reveals that
strongly anisotropic non-thermal phonon populations are established
throughout the first picoseconds of the dynamics.
A regime of thermal equilibrium is only re-established
by the ensuing anharmonic decay pathways (phonon-phonon coupling)
on timescales of the order of 50-100~ps.
To unravel the origin of the non-equilibrium lattice dynamics and its signatures in FEDS experiments,
we conduct first-principles calculations of the coupled electron-phonon dynamics based on the time-dependent Boltzmann formalism, whereby
electron-phonon and phonon-phonon scattering processes are explicitly accounted
for. Calculations of the structure factor further enable a direct comparison with the experimental data.
We show that the relaxation of excited electrons and holes is governed by the emission of
high-energy optical phonons within a restricted region of
the Brillouin zone, and it is responsible for the
anisotropic non-thermal phonon populations revealed by FEDS.

\section{Results and discussion}
\begin{figure*}[ht!]
\centering
\includegraphics[width=1\textwidth]{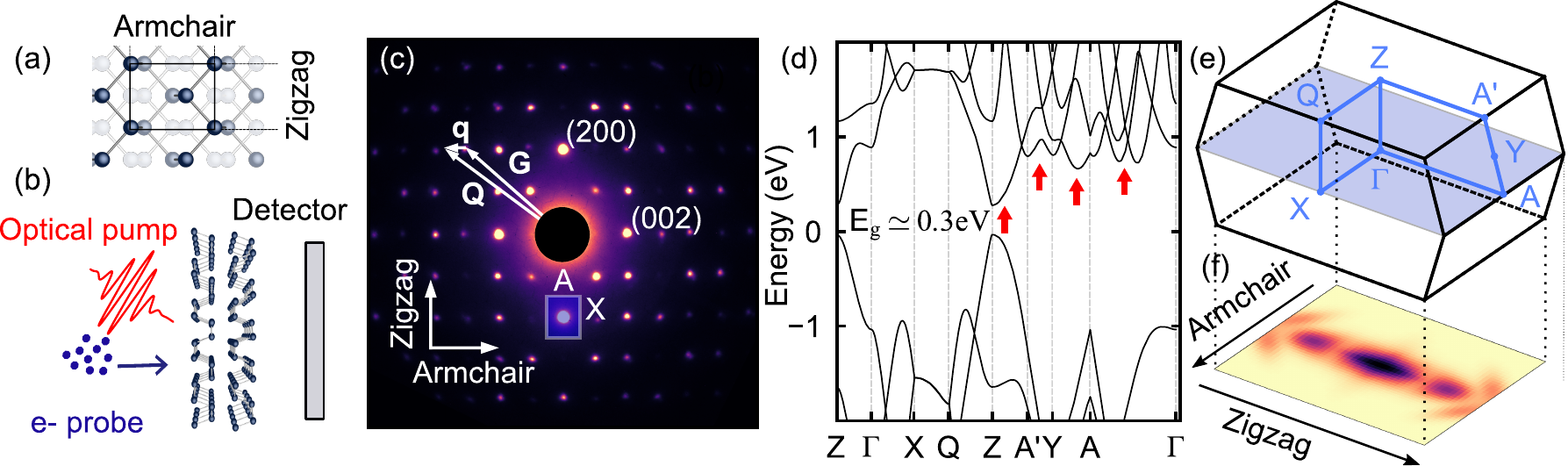}
\caption{(a) Top view of the BP crystal lattice. (b) Schematic illustration of
femtosecond electron diffuse scattering, with side view of the BP crystal lattice. (c) Representative transmission diffraction pattern
of BP. The Brillouin zone can be drawn around each Bragg peak, as
illustrated by the rectangle over the ($\overline{2}$00) reflection. An arbitrary
position in reciprocal space, $\bQ$, can always be expressed as $\bG + \bq$,
where $\bG$ is a reciprocal lattice vector and $\bq$ the phonon wavevector. 
(d) Electronic band structure as obtained from density-functional theory. 
A scissor shift of 0.2~eV has been applied to the conduction manifold 
to match the experimental band gap $\textrm{E}_\textrm{g} \simeq 0.3$~eV
\cite{Keyes1953}. (e) Brillouin zone and high-symmetry
points of BP. 
The blue shading marks the region of reciprocal 
space probed by our FEDS measurements.
(f) Momentum distribution of photo-excited carriers approximated by a
Fermi-Dirac function $f_{n\mathbf{k}}$ is shown (dark regions indicate more excited
carriers). The colored rectangles indicate phonons groups, see text.} 
\label{fig:1}
\end{figure*}

The layered orthorhombic crystal structure of BP is illustrated in Figure~\ref{fig:1}(a) and (b) from 
a top and side view, respectively, whereas its  Brillouin zone (BZ) 
and main high-symmetry points (labelled according to the convention of Ref.~\citenum{Ribeiro_2018}) 
are reported in  Figure~\ref{fig:1}(e). 
The equilibrium electron diffraction pattern of Figure~\ref{fig:1}(c) provides a direct 
view of the reciprocal lattice for momenta within the X-$\Gamma$-A plane in the BZ 
[shaded blue plane in Figure~\ref{fig:1}(e)]. Bright high-intensity features 
arise for transferred momenta matching the reciprocal lattice vectors ${\bf G}$, 
according to Bragg's law for the interference condition. 
These measurements are in good agreement with previous TEM experiments \cite{Gomez2014}.
Besides the pronounced anisotropy of the BP crystal lattice, 
marked by different structural motifs along the \textit{armchair} 
and \textit{zigzag} directions (Figure~\ref{fig:1}(a)), 
striking signatures of anisotropy also manifest themselves in the electronic properties. 

The electronic band structure, obtained from density-functional theory and  
illustrated in Figure~\ref{fig:1}(d), exhibits a direct gap at the Z point and 
a conduction band characterized by several local minima in the vicinity of the 
Y, A, and A$^\prime$ high-symmetry points.  
The local minima in the conduction band  thus involve crystal 
momenta with an in-plane component directed primarily along the zigzag direction. 
Conversely, no local minima arise in the conduction band 
along $\Gamma$-X and Z-Q (armchair direction).
The anisotropic character of the band structure is shown below to 
influence profoundly the non-equilibrium dynamics of electrons and 
phonons in BP, leading to the emergence of a striking anisotropy 
in the phonon population following the interaction with light pulses.  
\subsection{Femtosecond electron diffuse scattering measurements}
To investigate the non-equilibrium lattice dynamics of BP 
with momentum and time resolutions, we perform FEDS measurements 
on a free-standing thin film of BP. The sample has 
an estimated thickness of 39~$\pm$~5~nm and it has been  
obtained by mechanical exfoliation of a bulk crystal. 
In FEDS, a laser pulse is employed to drive the system into an 
excited electronic state. After a controllable time delay $t$, 
the sample is probed by an electron pulse which diffracts off the lattice. 
The distribution of the diffracted electrons generated by this procedure
provides a direct probe of the non-equilibrium dynamics of the lattice in 
reciprocal space \cite{2015Wald}. A schematic illustration of the experiment 
is reported in Figure~\ref{fig:1}(b). 
Here, the BP flake is photo-excited with a light pulse 
with energy $h\nu = 1.61$~eV and polarization aligned along the armchair direction. 
Additional measurements using a pump energy $h\nu =0.59$~eV 
are reported in the Supporting Information. 
All measurements are performed at 100~K. 
The initial density of photo-excited electrons and holes induced by the pump pulse 
is estimated to $n_e = 7.3 \cdot 10^{13}$ cm$^{-2}$ (see Supporting Information).

\begin{figure}[t]
\includegraphics[width=0.5\linewidth]{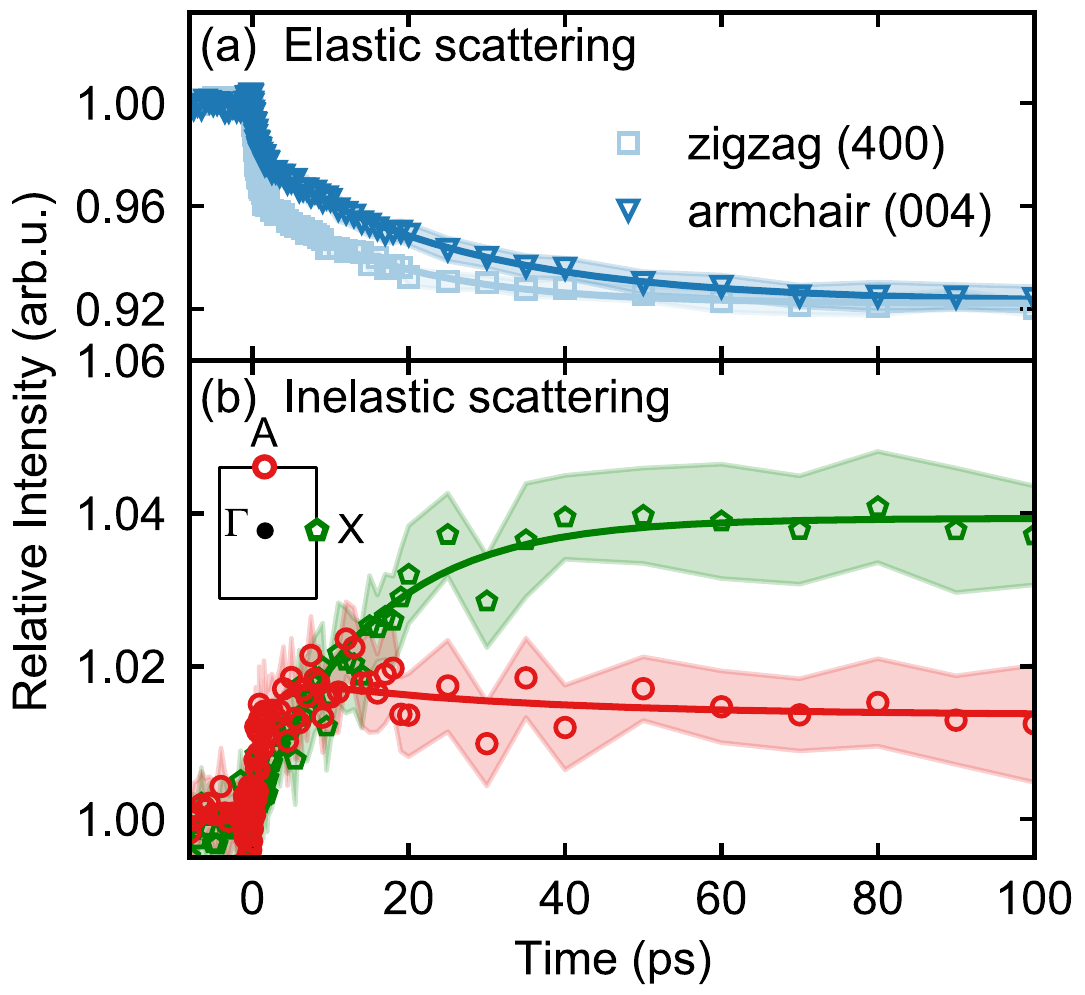} \caption{(a) Exemplary
anisotropic elastic scattering signals for \textit{zigzag} (squares) and
\textit{armchair} reflections (triangles).  (b) \deleted[]{inelastic}
{diffuse} scattering
signal at A (circles) and X (pentagons) around the (400) reflection. The data in
both panels is the average over the Friedel pair (e.g. (400) and
($\bar{4}$00)). The error estimates represent the standard error of the mean
signal over multiple delay scans.}
\label{fig:2}
\end{figure}

Figure~\ref{fig:2}(a) illustrates the relative 
intensity changes of the (400) and (004) Bragg peaks -- located along the zigzag 
(squares) and armchair (triangles) directions, respectively -- throughout the non-equilibrium dynamics 
of the lattice.
A clear fingerprint of anisotropic lattice dynamics is revealed by the different 
time dependence of these elastic scattering signals. 
The dynamics of both armchair and zigzag 
reflections are well-captured by bi-exponential decays,
with a fast time constant of 500~fs and a slower time constant 
of 20~ps. 
This behaviour was described in detail in Ref.~\cite{Zahn2020}, 
where some of us investigated the dynamics of the Bragg reflections in BP, 
revealing non-thermal phonon distributions persisting for tens of picoseconds.

To obtain a momentum-resolved picture of the non-equilibrium lattice dynamics of BP, we go beyond the analysis of the elastic scattering 
signals and we inspect the transient signatures of 
inelastic scattering processes as revealed by FEDS. The contribution of the different high-symmetry points to the FEDS intensity can be singled out by dividing the diffraction pattern into BZs around each Bragg reflection peak, as illustrated by the shaded rectangle in 
Figure~\ref{fig:1}~(c) for the ($\overline{2}$00) reflection. 
Exemplary time-resolved FEDS signals around the (400) Bragg peak are shown in Figure~\ref{fig:2}~(b) for the A (circles) and X (pentagons) points in the BZ. As diffuse scattering occurs primarily through phonon-induced scattering processes, 
the signal measured at a given point ${\bf q}$ in the BZ
reflects the phonon population with the same momentum \cite{Stern2018,
Waldecker2017, Trigo2010,Trigo2013, Wall2018, Cotret2019}. The red curve in
Figure~\ref{fig:2}~(b) indicates the relative intensity of the 
FEDS signal as a function of time at the A point. Similar dynamics
are observed at all the investigated A points. A bi-exponential fit to the data
yields a rising time constant of 1.7$\pm$0.1~ps, followed by a slower
relaxation of 30$\pm$2~ps. We note that the 1.7~ps time constant does not
appear in an elastic scattering analysis. 
The time evolution of the diffuse signal at X, shown in green in Figure~\ref{fig:2}~(b),  
reveals a drastically different phonon dynamics at the X point as compared to the A point. 
Fitting with an exponential function yields a time constant of 14.3~$\pm$~0.1 ps. 
These measurements indicate a striking anisotropy of the transient FEDS intensity in the 
BZ, suggesting a highly momentum-dependent excitation and relaxation of the lattice 
following photo-excitation. 

A comprehensive view of transient phonon
distributions in momentum space is shown in Figure~\ref{fig:3}~(a-c), at
pump-probe delays of 2, 10, and 50~ps. This set of data demonstrates profound changes in the diffuse scattering
signal as pump-probe delay increases, reflecting different phonon populations
at different times. While the diffuse pattern
at 2 ps is weak and displays faint lines in the $\Gamma-$A direction, the
diffuse signal at 50 ps is more pronounced, 
differently shaped, and more anisotropic. A closer inspection of the changes in
diffuse scattering signal around Bragg
reflections, shown as insets in Figure~\ref{fig:3}~(c), reveals high anisotropy
between the intensities along the two main crystal axes at 50 ps. These highly
anisotropic dynamics within a given BZ, and between BZs, highlight the value of
time-resolved diffuse scattering as direct
probes of transient non-thermal phonon distributions in momentum space
[Figure~\ref{fig:3}~(a-c)].

\subsection{Theoretical modeling of non-equilibrium lattice dynamics}
To gain further insight into the non-equilibrium dynamics of the lattice, 
we perform first-principles calculations of the coupled 
electron-phonon dynamics of BP based on the time-dependent 
Boltzmann equations \cite{caruso2021}: 
\begin{align}
\label{eq:bte_f}  
{\D_t f_{n\bk}(t)}     
&= \Gamma^{\rm ep}_{n{\bk}}[f_{n\bk}(t), n_{\bq\nu}(t)] \\
\label{eq:bte_n}  {\D_t n_{\bq \nu}(t)}  &= \Gamma^{\rm pe}_{{\bq\nu}}[f_{n\bk}(t), n_{\bq\nu}(t)]  + \Gamma^{\rm pp}_{{\bq\nu}}[n_{\bq\nu}(t)]
\end{align}
Here, ${\D_t} = \D / \D t$, $f_{n\bk}$ denotes the electron distribution function for  electron band index $n$ and electron momentum $\bk$, and $n_{\bq \nu}$ the phonon distribution function for wavevector $\bq$ and branch index $\nu$.
Equations~\eqref{eq:bte_f} and \eqref{eq:bte_n} account seamlessly for the
effects of electron-phonon and phonon-phonon scattering on the ultrafast dynamics 
of electrons and phonons with momentum resolution. 
The influence of the electron-phonon interaction on $f_{n\bk}$ ($n_{\bq \nu}$)
is accounted for by the collision integral $\Gamma^{\rm ep}_{n{\bk}}$ 
($ \Gamma^{\rm pe}_{{\bq\nu}}$), whereas the phonon-phonon scattering due to 
lattice anharmonicities is accounted for via $\Gamma^{\rm pp}_{{\bq\nu}}$.
In short, Eqs.~\eqref{eq:bte_f} and \eqref{eq:bte_n} have been solved for electron (phonon) momenta 
within the Q-Z-A$^\prime$ (X-$\Gamma$-A) plane in the BZ by time-stepping the time derivative with intervals of 1~fs for a total simulated time of 
100~ps (10$^5$ time steps), with the collision integrals being recomputed at each 
time step. A detailed account of the numerical implementation and explicit expressions 
for the collision integrals have been reported elsewhere.\cite{caruso2021} 

As initial condition for the time propagation,
we consider an electronic excited state characterized by a density 
$n$ of electrons and holes excited to the conduction and valence bands, respectively. 
This state is realized by defining electronic occupations 
in the conduction band according to 
$f_{n\bk}^{0} (\mu_e,T_{\rm el}^{0}) = [e^{(\ve_{n\bk}
-\mu_e/k_{\rm B}T_{\rm el}^{0}}+1 ]^{-1}$. 
$\mu_{\rm e}$ is the  chemical potential of the electrons in the conduction band,
which is obtained by solving the integral equation 
$n= {\Omega_{\rm BZ}^{-1}} \sum_m^{\rm cond.} \int {d\bk} f^0_{m\bk} ( \mu_{\rm e} , T_{\rm el}^{0})$, where
$n= 7.3\cdot10^{13}$~cm$^{-2}$ is the density of photo-excited carriers estimated in the experiment (see Supporting Information). 
A similar treatment is applied to holes in the valence band. 
The initial electronic temperature $T_{\rm el}^{0}=7000$ K 
is related to the excess energy of the excited electrons and holes, and  
it is chosen such that the final vibrational temperature of the lattice 
after thermalization $T^{\rm vib}_{\rm fin}$ matches the experimental estimate 
of $300$~K. The lattice is initially at thermal equilibrium, 
with phonon occupations defined according to 
the Bose-Einstein statistics $n_{\bq\nu}^{\rm BE} = [e^{\hbar\w_{\bq\nu} /k_{\rm B}T}-1 ]^{-1}$,
at the same temperature of experiments $T=100$~K. 

\begin{figure}[t]
\includegraphics[width=1\textwidth]{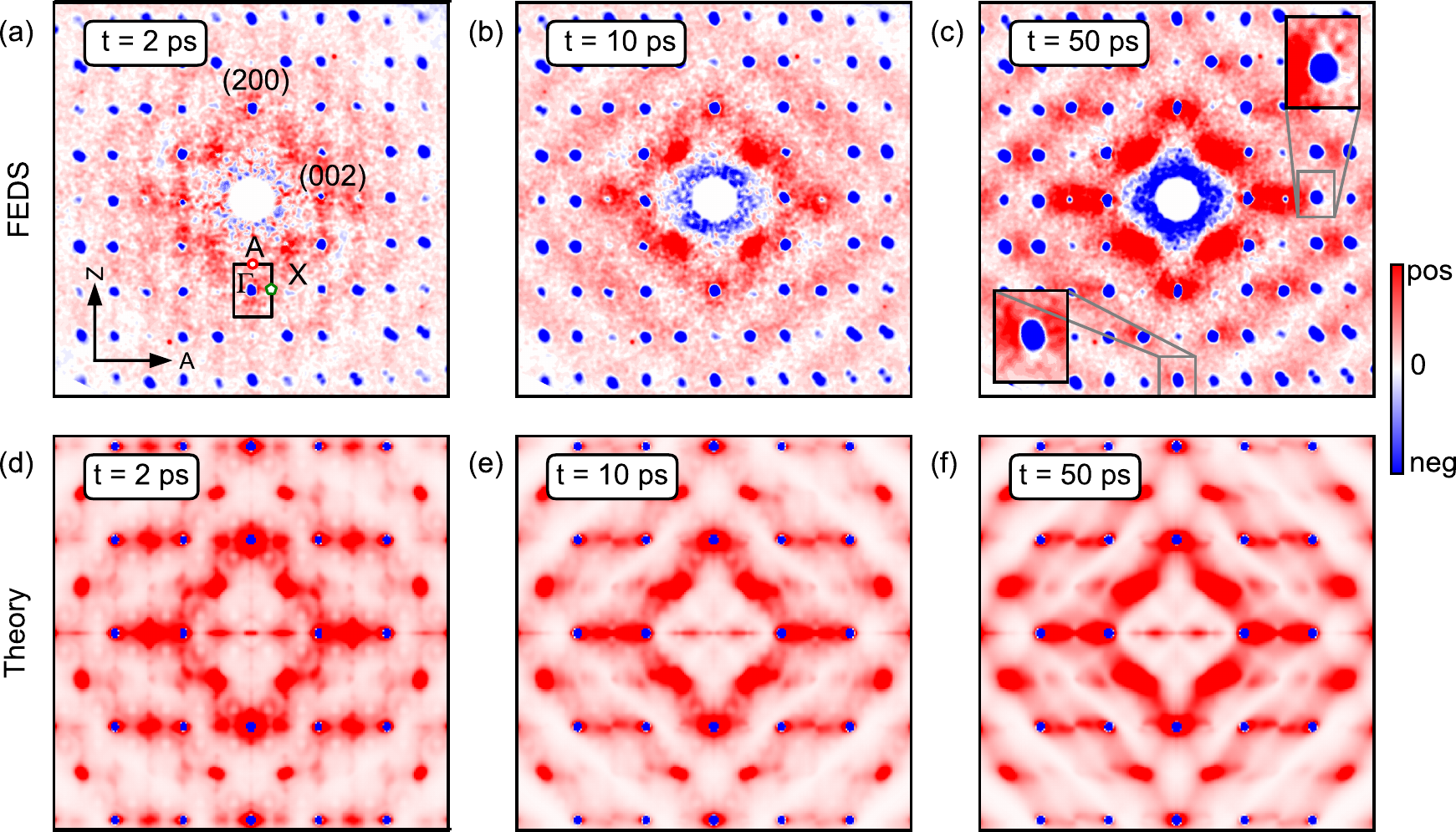}
\caption{(a-c) Momentum-resolved electron diffraction signals, $I(\mathbf{Q}, t)- I(\mathbf{Q}, t \leq t_{0})$, at pump-probe delays of 2 ps, 10 ps, and 50 ps.
Two-fold symmetrized data \cite{RendeCotret2018}, raw data shown in
Supporting Information. The Bragg reflections (blue dots) are negative
due to the Debye-Waller effect. The \deleted[]{inelastic} {diffuse}
background (red) qualitatively evolves as a function of pump-probe delay.
Selected Brillouin zones are shown in inset for the (004) and the
($\overline{4}00$) reflections on the 50 ps map. All data are normalized to a
common number. (d-f) Simulated non-equilibrium scattering signals at pump-probe
delays of 2 ps, 10 ps, and 50 ps. The phonon temperatures are based on the
non-thermal model described in the text and shown in Figure 4 (a). All data are
normalized to a common number.} \label{fig:3}
\end{figure}

\begin{figure*}[t]
\includegraphics[width=0.98\textwidth]{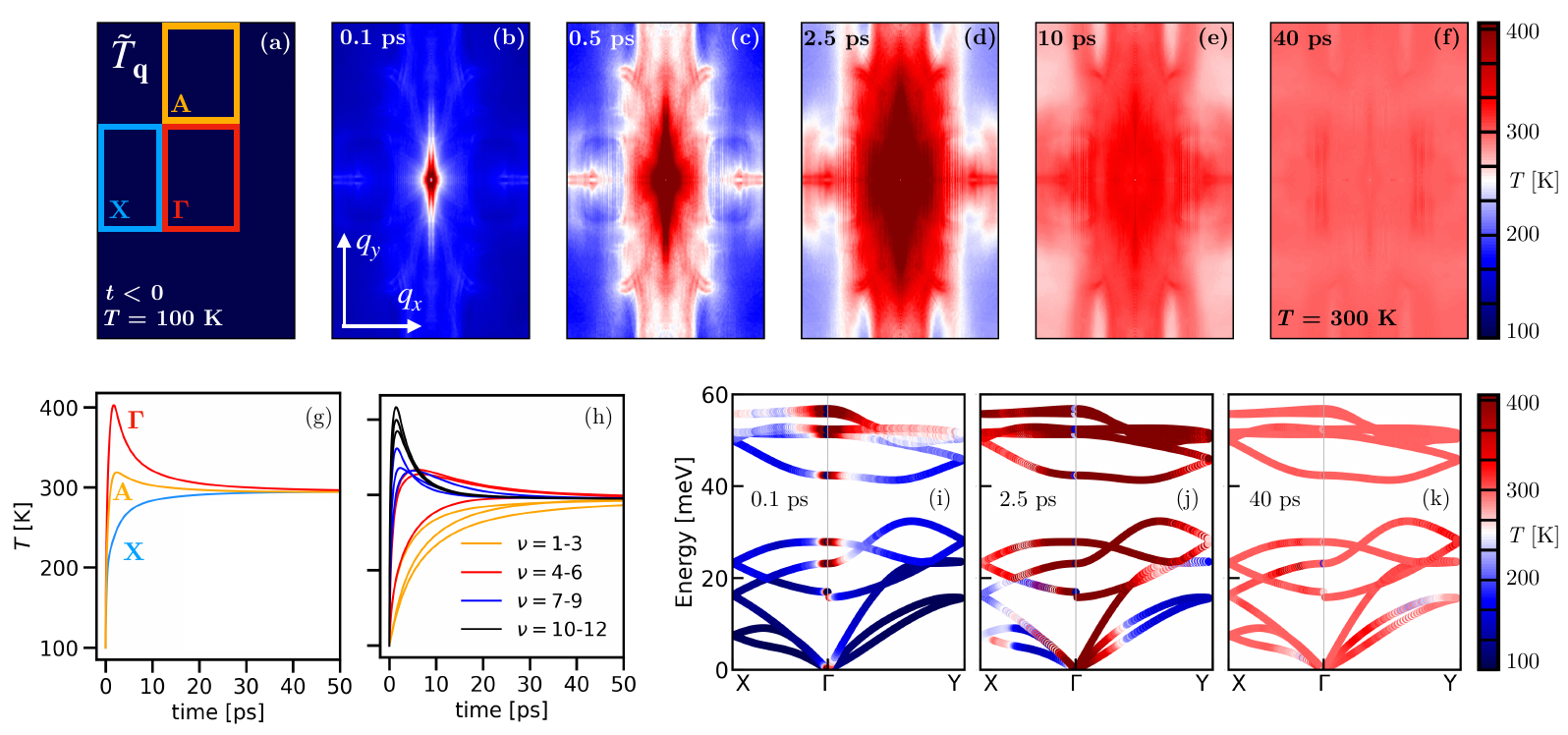}
\caption{
(a) Effective vibrational temperature $\tilde T_{\bf q}$ 
for crystal momenta in the X-$\Gamma$-A plane of the Brillouin zone 
before excitation ($t<0$), and at several time delays throughout the 
non-equilibrium dynamics of the lattice (b-f). 
(g)~Time-dependence of $\tilde T_{\bf q}$  for 
momenta around the high-symmetry points $\Gamma$ (red), A (yellow), and X (blue). 
Each curve has been obtained by averaging  $\tilde T_{\bf q}$ for momenta within the 
regions highlighted in (a) at each time step. 
(h) Time-dependence of the branch-resolved  vibrational temperature $T_{\nu}$ (averaged over momentum).  
$\nu=1-3$ denote the acoustic branches, $\nu=10-12$ the highest-energy optical phonons, etc. 
(i-k) Branch and momentum resolved effective vibrational temperatures, 
superimposed to the phonon dispersion as a color coding, for $t=0.1$ (i), 2.5 (j), and 40~ps (k). 
}
\label{fig:4}
\end{figure*}

From the phonon distribution function $n_{\bq\nu}(t)$, we can calculate the momentum-resolved effective 
vibrational temperature of the lattice $\tilde T_{\bq}=N_{\rm ph}^{-1} \sum_\nu T_{\bq\nu}$, where $ N_{\rm ph}=12$ is the number of phonon branches of BP and 
$T_{\bq\nu}= \hbar \omega_{\bq\nu} \{k_{\rm B} \ln [ 1 + n_{\bq\nu}(t) ]\}^{-1}$. 
In Figure~\ref{fig:4}(a)-(f), we report $\tilde T_{\bq}$
at different time steps of the coupled electron-phonon dynamics for 
crystal momenta within the X-$\Gamma$-A plane of the BZ (shaded blue in 
Figure~\ref{fig:1}(e), corresponding to the plane probed in the FEDS experiments).  
Before excitation ($t<0$), the constant temperature $\tilde{T}_{\bq} = 100$~K in the BZ reflects 
thermal equilibrium. At $t=0.1$~ps, red features in Figure~\ref{fig:4}(b) 
indicate the enhancement in the phonon population around $\Gamma$ (zone center) 
and along the $\Gamma$-A high-symmetry line.
This anisotropy becomes more pronounced at later times, as shown in 
 Figure~\ref{fig:4}(c) and (d) for $t=0.5$ and 2.5~ps, respectively.  
As anticipated above, the origin of this behaviour is related to the 
anisotropy of the valence and conduction bands. 

Owing to the absence of local minima in conduction  
band along the armchair direction (i.e., $\Gamma$-X and Z-Q), 
the photo-excited electrons are constrained to occupy 
states with crystal momenta along the zigzag direction, i.e., 
where the available local minima are located (arrows in Figure~\ref{fig:1}~(d)).
This scenario is illustrated by  highly-anisotropic electronic occupations 
$f_{n{\bf k}}^0$ in the conduction band, reported in Figure~\ref{fig:1}~(f) 
for the initial electronic excited state defined above, 
arising from the partial filling of the available low-energy states.  
Due to momentum conservation, the relaxation of carriers to
the conduction band minimum (Z point) thus entails a predominant
emission of phonons with momenta ${\bf q}$ along the $\Gamma$-A directions. 
Based on this picture, the anisotropic increase of the vibrational temperature 
in the BZ reflects the phase-space constraints in the electron-phonon 
interactions, and thus in the relaxation path of photo-excited electrons and holes.
For $t=10$~ps  (Figure~\ref{fig:4}~(d)), the anisotropy of the vibrational temperature 
in the BZ is significantly reduced. On these timescales, 
phonon-phonon scattering -- accounted for via $\Gamma^{\rm pp}$ in Eq.~\eqref{eq:bte_n} -- 
counteracts the effects of the electron-phonon scattering by 
driving the lattice towards thermal equilibrium. For $t=40$~ps  (Figure~\ref{fig:4}~(d)),
thermal equilibrium is re-established at the temperature $T_{\rm vib}^{\rm fin}=300$~K. 

To gain further insight into the anisotropy of the lattice dynamics, 
we illustrate in Figure~\ref{fig:4}(g) the time dependence of the vibrational 
temperature $\tilde T_{\bf q}$ around the  X, $\Gamma$, and A regions
(obtained by averaging $\tilde T_{\bf q}$ over the rectangles in Figure~\ref{fig:4}(a))
throughout the first 50~ps of the dynamics.
For momenta around $\Gamma$ and A, the temperature reaches a maximum at 1.7 and
2.3~ps, respectively, whereas no maximum is observed around X. These timescales 
indicate the time required for the electrons to transfer energy to the lattice 
via electron-phonon scattering. 
The good agreement with the experimental time constant of 1.7~ps extracted from 
the rise of the FEDS intensity at $A$ (Figure~\ref{fig:2}(b)), 
suggests that transient changes of the FEDS intensities 
for timescales smaller than 2~ps reflect primarily the energy transfer 
from the electrons to the lattice driven by the electron-phonon coupling. 

In Figure~\ref{fig:4}(h), we report the average vibrational temperature 
for each phonon branch ($\tilde T_{\nu}=  \Omega_{\rm BZ}^{-1}\int d{\bf q}\, T_{\bq\nu}$) 
throughout the first 50~ps, whereas the vibrational temperatures superimposed to the phonon dispersion 
is illustrated in Figure~\ref{fig:4}(i-k). Because the contribution of each phonon mode to 
the carrier relaxation is dictated by its coupling strength, modes characterized 
by stronger coupling provide a preferential decay channel for the excited 
electrons and thus exhibit a higher vibrational temperature throughout 
the initial stages of the dynamics. In particular, Figures~\ref{fig:4}(h)-(k) indicate that 
the electron relaxation is dominated by the high-energy optical phonons, whereby 
the out-of-phase vibration of P atoms in the lattice leads to the largest electron-phonon coupling strength. 

%

To inspect directly the influence of the non-equilibrium lattice dynamics on the 
scattering intensity probed in the FEDS experiments, we conduct first-principles 
calculations of the structure factor by explicitly accounting for the influence of  electron-phonon interactions and anisotropic population of the vibrational modes in the 
unit cell. Specifically, we perform computations of the \textit{all-phonon} structure factor $I_{\rm all}(\bQ,T)$ \cite{Zacharias2021, Zacharias2021_2}. Indeed we find that taking into account multi-phonon effects is essential for an accurate reproduction of the experimentally observed diffraction patterns of BP seen in Figures \ref{fig:3}(a-c). The expression for $I_{\rm all}(\bQ,T)$ reads:
\begin{eqnarray}\label{all-phonon}
I_{\rm all}(\bQ,T) = N_p \sum_{\k \k'} f_\k (\bQ) f^*_{\k'} (\bQ)
       e^{-W_{\k\k'}(\bQ,T)}  \sum_{p} e^{ i \bQ \cdot [\bR_p + \bt_\k - \bt_{\k'} ] }  \, e^{ P_{p,\k\k'} (\bQ,T)}.
\end{eqnarray}
Here $N_p$ is the number of  $\bq$-points used to sample the first Brillouin Zone, $f_\k (\bQ)$ denotes the scattering amplitude of atom $\k$,  $W_{\k \k'}(\bQ,T)$ is the Debye-Waller factor, $\bt_\k$ represents the atomic positions and $\bR_p$ defines the position vector of unit cell $p$ contained in a Born-von K\'arm\'an supercell.
The phononic factor, $e^{ P_{p,\k\k'} (\bQ,T)}$, includes all orders of phonon processes and its exponent is given by:
\begin{eqnarray}\label{eqa1.8_b}
 P_{p,\k\k'} (\bQ,T)  &=& \frac{M_0 N^{-1}_p}{\sqrt{M_\k M_{\k'}}} \sum_{\bq \nu }   \< u^2_{\bq \nu} \>_T \,
\text{Re}\Big[ \bQ \cdot {\bf e}_{\k,\nu} (\bq) \bQ \cdot {\bf e}^{*}_{\k',\nu} (\bq) e^{i\bq \cdot {\bf R}_p} \Big], 
\end{eqnarray}
where $M_\k$ and $M_0$ are the atomic and reference masses, and the phonons are described by the eigenmodes ${\bf e}_{\k,\nu}(\bq)$ and frequencies $\omega_{\bq \nu}$.  A key quantity entering the equation of the structure factor is the mean-squared displacement of the atoms due to mode
$\bq  \nu$, defined as 
$\<u^2_{\bq\nu}\>_T= \hbar/(2M_0\omega_{\bq \nu})[2n_{\bq  \nu}(T) + 1 ]$.
The time-dependence of the all-phonon structure factor is encoded in
$\<u^2_{\bq\nu}\>_T$, which is directly related to phonon populations
$n_{\bq\nu}(T)$. 
To account for the influence of the non-equilibrium lattice dynamics on the
FEDS intensity, we evaluated Eq. \eqref{all-phonon} at
each time snapshot by populating phonons according to the vibrational temperatures
obtained from the solution of the time-dependent Boltzmann equation (Figure~\ref{fig:4}).

The calculated (non-equilibrium) all-phonon structure factor is shown in Figure~\ref{fig:3}~(d) for $t=2$~ps.
The intensity is relative to equilibrium at 100~K.
The calculation agrees well with the experimental FEDS intensity reported in Figure~\ref{fig:3}~(a) and 
it reproduces the main fingerprints of non-equilibrium lattice dynamics. 
In particular, 
the faint vertical high-intensity features which connect the Bragg peaks across different BZ 
-- and constitute a striking manifestation of the non-equilibrium state of the lattice -- 
are well captured by the simulations. 
The time dependence of the vibrational temperature in the BZ, illustrated in Figure~\ref{fig:4}, 
enable us to attribute these features to the higher population of 
phonons along the $\Gamma$-A direction which, in turn, arises
from the primary role played by these phonons in the relaxation 
of the excited electronic distribution. 
%
The calculated FEDS intensities at 10~ps and 
50~ps, shown in Figure~\ref{fig:3}~(b) and (c), respectively, further capture 
the emergence of a diamond-shaped diffraction pattern that characterises the 
return to thermal equilibrium.

\begin{figure}[t]
\includegraphics[width=0.48\textwidth]{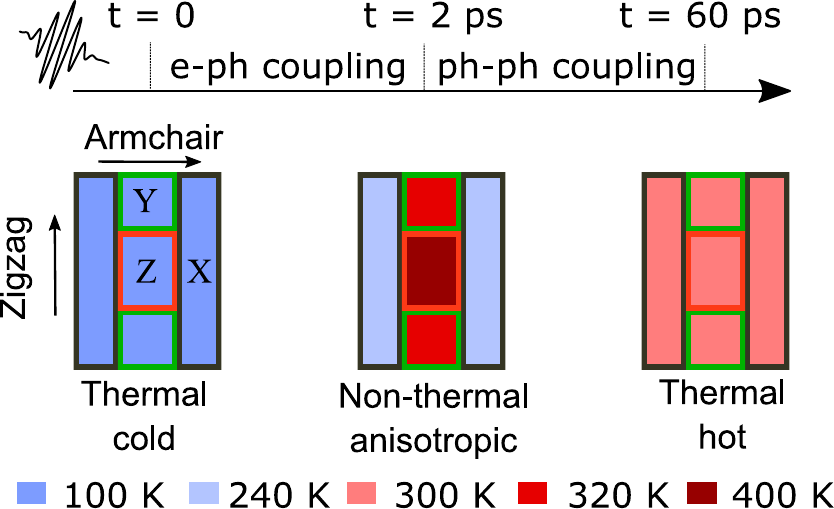}
\caption{Sketch of the non-equilibrium dynamics and thermalization 
of the BP lattice following photo-excitation.} \label{fig:5}
\end{figure}

These findings enable us to establish the picture sketched in Figure~\ref{fig:5}
for the non-equilibrium dynamics and thermalization of vibrational degrees of freedom in BP: 
After the creation of an excited electronic distribution  
by a laser pulse, electrons (holes) in the conduction (valence) band undergo electron-electron 
scattering and occupy the band edges according to Fermi-Dirac statistics. 
This results into a highly anisotropic distribution of
photo-excited carriers in the BZ, predominantly populating the Z, Y, A, and A$^\prime$
pockets. 
Within 2~ps after photo-excitation, electrons and holes lose their excess energy upon emitting phonons. 
Momentum selectivity in the phonon emission leads to the primary excitation of 
phonons with momenta along the zigzag direction of the crystal, 
driving the lattice into a non-equilibrium regime characterized by a highly-anisotropic 
phonon population in the BZ (Figure~\ref{fig:4}(b-d)). 
Distinctive fingerprints of this regime are visible in  
the FEDS intensity at $t=2$~ps (Figure~\ref{fig:3}(b)). 
%
%
The ensuing {\it hot}-phonon population subsequently thermalizes with other
lattice vibrations via phonon-phonon scattering, 
thereby driving the lattice towards thermal equilibrium (i.e., $T_{\bq\nu} =$~const.) within 50~ps, 
and leading to the thermalized FEDS intensity reported in Figure~\ref{fig:3}(c). 

\section{Conclusions}
We have provided a comprehensive picture of the microscopic energy flows in the crystal lattice of BP following photo-excitation of the electrons. The time- and momentum-resolved {diffuse} 
scattering experiments have revealed that highly-anisotropic 
transient phonon populations are established upon photo-excitation.  
By accounting explicitly for electron-phonon and phonon-phonon scattering 
within an ab-initio theoretical description of the coupled electron-phonon 
dynamics, we have demonstrated that this behaviour can be attributed to the 
preferential emission of high energy optical phonons along the zigzag direction of the BP 
lattice throughout the relaxation of the photo-excited electronic distribution. 
This picture is corroborated by the good agreement between the calculated 
all-phonon structure factors and the measured  FEDS intensity throughout 
the different stages of the non-equilibrium dynamics of the lattice. Our approach can be extended to 2D materials, and could be exploited more broadly in many areas of material sciences and condensed matter physics, ranging from transport to superconductivity phenomena. For instance, it could be employed to reveal energy transfer pathways accross interfaces in van der Waals heterostructures, or to identify specific phonons involved in complex electron-phonon dynamical processes such as polaron formation or phase transitions.
%
\section{Methods}
\subsection{Sample preparation and thickness determination}
The thin black phosphorus (BP) flake was obtained by standard mechanical exfoliation performed in air. The samples were then quickly imaged in the optical microscope and subsequently transferred to a load-lock chamber connected to our main experimental chamber in ultra-high vacuum. We estimate the total exposure to air to be less than one hour. We found that this method yielded diffraction patterns consistent with previous experimental works \cite{Gomez2014}. Given the multilayer nature of the samples (40 nm corresponds to roughly 80 layers), the observed scattering signals predominantly arise from the bulk as opposed to the oxidized surface layers of the flake. We note, however, the presence of forbidden reflections (h + l = 2n + 1) in the diffraction patterns. Such forbidden reflections were also observed in previous works, but their origin could not be attributed with certainty \cite{Gomez2014}. We postulate that they may be caused by stacking faults or structural deviations at the surface. These additional reflections do not alter the overall agreement between experiment and theory.  \par
The flake thickness was estimated by transmission measurements in an optical microscope in combination with transfer matrix calculations and the optical constants of BP \cite{Jiang2018}.
\subsection{Computational details}
First-principles calculations employed the primitive cell of bulk BP (point group D$_{2h}$ and space group Cmce) that contains 4 atoms~\cite{Ribeiro_2018}. All calculations were performed using the PBE generalized gradient approximation~\cite{GGA_Pedrew_1996} to density functional theory. We employed planewaves basis sets and  Troullier-Martins norm-conserving pseudopotentials~\cite{Troullier_Martins_1991} as implemented in the
{\tt Quantum ESPRESSO} suite~\cite{QE}. The planewaves kinetic energy cutoff was set to 90~Ry and the sampling of the Brillouin zone  was performed using a uniform 12$\times$10$\times$10 ${\bf k}$-point grid. We determine the interatomic force constants by means of density-functional perturbation theory calculations~\cite{Baroni_2001}  using a $5\!\times\!5\!\times\!5$ Brillouin-zone ${\bf q}$-grid. 
The full set of phonon eigenmodes and eigenfrequencies was obtained by using standard Fourier interpolation of dynamical matrices on a 50$\times$50$\times$50 ${\bf q}$-point grid. Such a dense grid guarantees a fine resolution of the calculated structure factor maps. The phonon band structure over a chosen high-symmetry path is shown in Figure S2. 

\par
\begin{acknowledgement}
This work was funded by the Max Planck Society, the European Research Council
(ERC) under the European Union’s Horizon 2020 research and innovation program
(Grant Agreement Number ERC-2015-CoG-682843), and partially by the Deutsche
Forschungsgemeinschaft (DFG) - Projektnummer 182087777 - SFB 951.
F.C. acknoledges funding by the DFG -- Projektnummer 443988403.
H.S.~acknowledges support by the Swiss National Science Foundation under Grant
No.~P2SKP2\textunderscore184100. 
M.Z. acknowledges financial support from the Research Unit of Nanostructured Materials Systems (RUNMS) and program META$\Delta$I$\Delta$AKT$\Omega$P.
Y.Q.~acknowledges support by the Sino-German
(CSC-DAAD) Postdoc Scholarship Program (Grant No.~57343410). {Y.W.W
acknowledges funding from the DFG within the Emmy Noether program under Grant
No. RE 3977/1}. We thank Maciej Dendzik for helpful discussions, and Laurent
Ren\'e de Cotret for his open-source software. F.C.~and C.D.~acknowledge Dino
Novko for useful discussions. 
\end{acknowledgement}

\begin{suppinfo}
Estimation of excited carrier density; Diffuse scattering maps: raw data and influence of pump photon energy
\end{suppinfo}

\bibliographystyle{achemso}
\bibliography{bibliography}

\end{document}


\flushbottom
\maketitle

\thispagestyle{empty}
\newpage
\section{Estimation of excited carrier density}
The incident fluence on the BP flake is $I_{\textrm{inc}}$ = 98 J/m$^2$. \added[]{The pump was polarized along the \textit{armchair} direction, determined by rotating a waveplate in the pump arm and maximizing the Debye-Waller effect at 50 ps (maximum absorption). Given the highly anisotropic optical absorption of BP, specifying the pump polarization is essential to employ the proper value of refractive index, which matters for the estimation of excited carried density.} The complex refractive index of BP along the \textit{armchair} direction at 800 nm is 3.19 + 0.29i, as estimated from a previous work \cite{Jiang2018}. We use transfer matrices to calculate the transmitted part, T, as well as the reflected part, R, of the incident fluence. This yields the absorbed fluence  $I_{\textrm{abs}} = (1 - R - T) \cdot I_{\textrm{inc}}$. The carrier density per square centimeter is then:

\begin{equation*}
\deleted[]{
n_{\mathrm{e}} = 10^{-6}\cdot  \frac{1}{E_{\textrm{ph}}} \cdot \frac{I_{\textrm{abs}}}{d} = 1.42 \cdot 10^{21} \quad \textrm{electrons}/\textrm{cm}^3,}
\added[]{
n = 10^{-6}\cdot  \frac{1}{E_{\textrm{ph}}} \cdot \frac{I_{\textrm{abs}}}{d} \cdot c \cdot \frac{1}{2}= (7.3 \pm 0.9) \cdot 10^{13} \quad \textrm{electrons}/\textrm{cm}^2,}
\end{equation*}

where $E_{\textrm{ph}}$ is the energy of one pump photon, $d$ is the flake thickness, which we estimate to 39 $\pm$ 5 nm, $c = 10.46 \cdot 10^{-10}$ m is the unit cell length in the out-of-plane direction and the factor 1/2 accounts for the two layers per unit cell.

\newpage
\section{Diffuse scattering maps: raw data and influence of pump photon energy}
In supplementary Figure \ref{fig:tile} we show a series of \deleted[]{inelastic} \added[]{diffuse} scattering maps at chosen delays and for different pump photon energies and/or data processing steps. Panels (a-c) and (j-l) correspond to the experimental and theoretical results shown in Figure 3 of the main text, respectively. They are reproduced here for convenience. Panels (d-f) display the \deleted[]{inelastic} \added[]{diffuse} scattering signals obtained when pumping the sample with 0.59 eV photons. The same initial excited electron density is assumed for both the 0.59 eV and 1.61 eV experiments, based on the Bragg peaks' reduction at 300 ps. While the \deleted[]{inelastic} \added[]{diffuse} scattering signatures at 2 ps seem slightly more intense for the 1.61 eV pump, we observe the same qualitative thermalization process in both cases. Since the picosecond dynamics does not seem to depend significantly on the initial carrier distribution after excitation, the assumption of a thermalized electron system for our purposes seems justified. In the main text and in panels (a-f), we show two-fold symmetrized data. This processed data is obtained using the open-source Python environment developed by De Cotret et al. \cite{RendeCotret2018}. Prior to symmetrization, it was verified that the time-dependence of the elastic scattering for Friedel pairs showed the same dynamics within error margin. The raw data for the 1.61 eV pump are shown in panels (g-i). \added[]{Comparing panels (a-c) and (g-i), we observe peak splitting effects at large scattering vectors for the symmetrized data. This artifact arises from field distortions of our magnetic lens, which contains aberrations at higher scattering vectors.}
\newpage
\section{Supplementary Figures}

\begin{figure}[th!]
\renewcommand{\figurename}{Supplementary Figure}
\centering
\includegraphics[width=\linewidth]{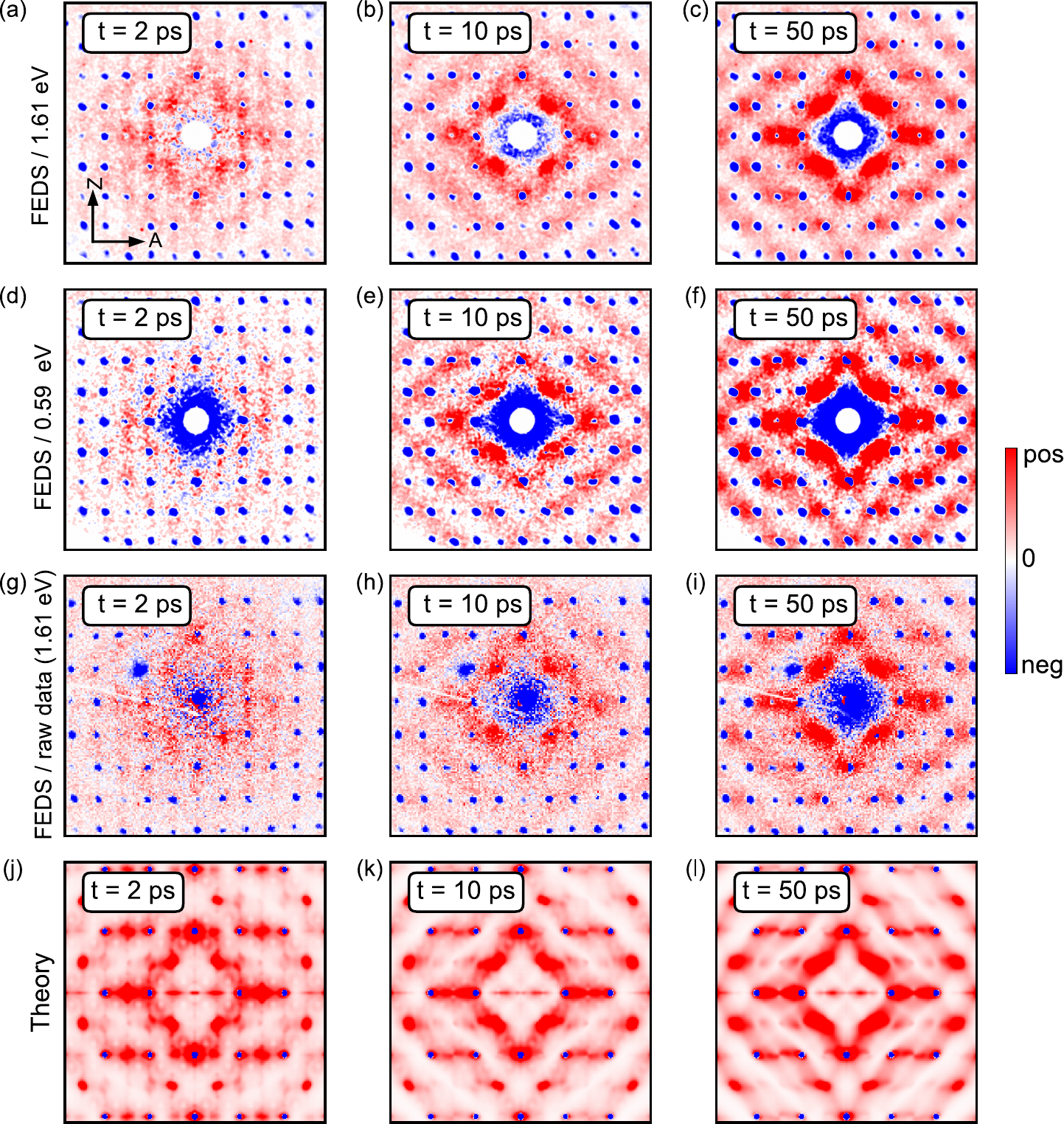}
\caption{\deleted[]{Inelastic} \added[]{Diffuse} scattering maps at 2 ps, 10 ps, and 50 ps for different pump photon energies and data processing steps. (a-c) Two-fold symmetrized, 1.61 eV pump. (d-f) Two-fold symmetrized, 0.59 eV pump. (g-i) Raw data (binned), 1.67 eV pump. (j-l) Theory results obtained assuming an excited carried density of 1.42 $\cdot 10^{21}$ electrons/cm$^3$.}
\label{fig:tile}
\end{figure}

\newpage

\begin{figure}[th!]
\renewcommand{\figurename}{Supplementary Figure}
\centering
\includegraphics[width=0.8\linewidth]{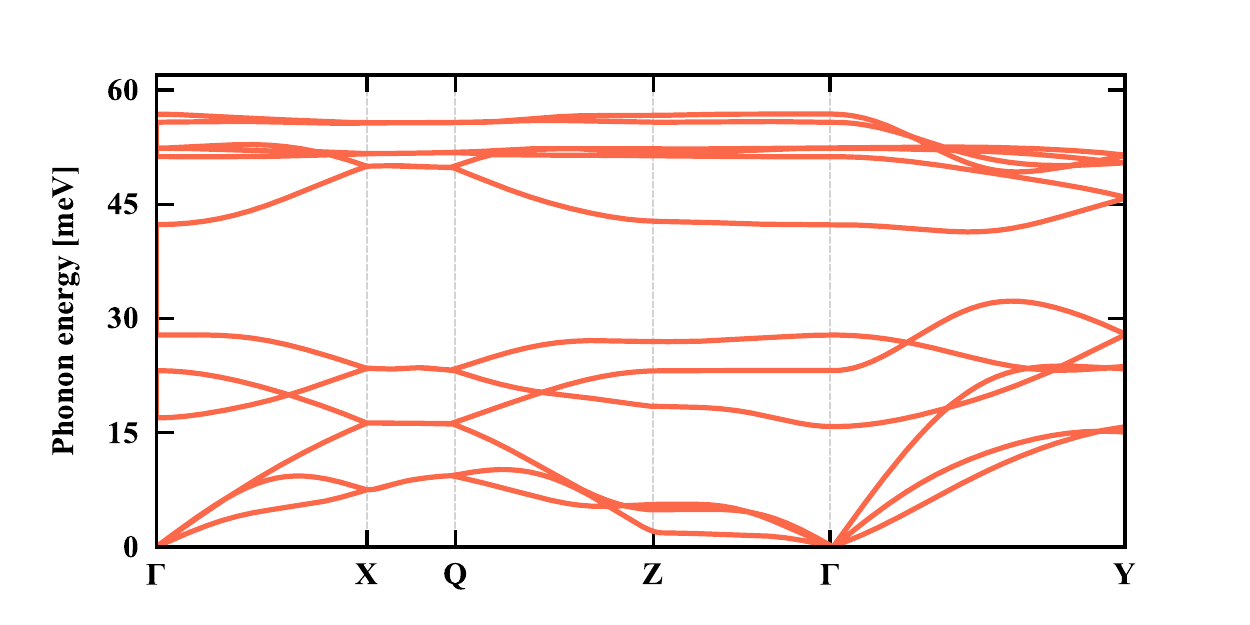}
\caption{Phonon dispersion of BP calculated from density-functional perturbation theory over the $\Gamma$-X-Q-Z-$\Gamma$-Y path.}
\label{fig:ph_band_structure}
\end{figure}





\bibliography{bibliography}